\documentstyle[aps,prb,epsfig]{revtex}
\begin{document}
\baselineskip=0.5cm
\renewcommand{\thefigure}{\arabic{figure}}
\title{Damping of long-wavelength collective excitations in quasi-onedimensional Fermi liquids}
\author{F. Capurro, M. Polini, M. P. Tosi$^*$}
\address{INFM and Classe di Scienze, Scuola Normale Superiore, I-56126 Pisa, Italy
}
\maketitle
\vspace{1.5 cm}

\noindent {\bf Abstract}
\vspace{0.2 cm}

The imaginary part of the exchange-correlation kernel in the longitudinal current-current 
response function of a quasi-onedimensional Fermi liquid is evaluated by an approximate 
decoupling in the equation of motion for the current density, which accounts for processes of 
excitation of two particle-hole pairs. The two-pair spectrum determines the intrinsic damping rate 
of long-wavelength collective density fluctuations, which is calculated and contrasted with a 
result previously obtained for a clean Luttinger liquid.
\vspace{0.4 cm}


\vspace{3 cm}

\noindent $^*$Corresponding author; e-mail: tosim@sns.it

\newpage

\section*{1. Introduction}
An established theoretical viewpoint on one-dimensional (1D) fermion fluids is that the 
fermion-fermion interactions induce a Luttinger-liquid state, in which the momentum distribution 
does not have a discontinuity across the Fermi surface and spin-charge separation occurs [1]. A 
quasi-1D system which is of special interest both for basic physics and for technology is 
obtained by confining electrons in GaAs-based semiconductor quantum wire structures. The 
excitation spectrum of such a system in the 1D quantum limit, where only the lowest subband is 
populated, has been measured by Go${\tilde {\rm n}}$i {\it et al.}~[2] by resonant inelastic light scattering. Focusing 
here on the intrasubband transitions, they consist of charge-density and spin-density collective 
excitations as well as of single-particle (electron-hole pair) excitations. As discussed by Wang et 
al. [3], no definitive evidence for Luttinger-liquid {\it versus} Fermi-liquid behaviour has been 
obtained from these data. The Fermi surface can be restored in a weakly interacting 1D system 
by scattering of the electrons against impurities or thermal fluctuations [4] and at any rate the 
differences between a Luttinger liquid and a Fermi liquid are most marked in the one-electron 
spectrum, rather than in the particle-hole spectrum studied by Go${\tilde {\rm n}}$i {\it et al.}~[2]. In particular, the 
plasmon dispersion relation that these authors have measured is well described by the standard 
random-phase (RPA) expression [5], in a regime of wave number and wire thickness where the 
logarithmic factor from long-range Coulomb interactions is slowly varying so that an almost 
linear dispersion is observed. 

A number of authors have recently drawn attention to the possible relevance of the 
Tomonaga-Luttinger model in determining the density profile and the low-lying excitations of 
quasi-1D quantum gases of atoms inside highly anisotropic traps [6 - 8]. The extremely high 
purity and low temperature of these systems permit refined experimental studies using the tools 
of atomic physics and quantum optics. With regard to excitations in a clean Luttinger liquid, 
their lifetime has been calculated by Samokhin [9] in the case of short-range (delta-function) 
interactions. After stressing that a linear dispersion relation, for which the laws of energy and 
momentum conservation are simultaneously satisfied, leads in second-order perturbation theory 
to a divergent decay rate in the limit of an infinitely long sample, Samokhin carried out a self-
consistent calculation of the damping of excitations from their mutual collisions that are allowed 
by a finite band curvature [10]. 

In the above perspective it seems useful to carry out a similar calculation for the lifetime of 
collective excitations in a 1D Fermi liquid and to contrast the result with that obtained by 
Samokhin [9] for a 1D Luttinger liquid. As is known from the study of interacting electron 
fluids in higher dimensionalities [11 - 13], the damping of a collective excitation at long 
wavelengths is determined by its decay into two particle-hole pairs and the rate of this process 
can be evaluated from the imaginary part of the exchange-correlation kernel entering the 
equation of motion for the current density. The same approach applied to a neutral Bose fluid 
describes the Beliaev-Popov mechanism for the damping of phonons [14] and more generally 
allows the evaluation of its visco-elastic spectra [15]. The only important conceptual difference 
for the quasi-1D liquid is that transverse currents are quenched in this case by the tight 
transverse confinement.  For a quasi-1D Fermi liquid this damping mechanism has only been 
studied in the asymptotic high-frequency limit [16]. 

The paper is organized as follows. Section~2 explicitly shows how the inclusion of band 
curvature in a non-perturbative way leads to the same equations of motion for the particle density 
and current density in a quasi-1D Fermi fluid as for Fermi fluids in higher dimensionalities. We 
proceed from there to evaluate the exchange-correlation kernel at T = 0 in Section~3 and the 
related damping rate of collective excitations in Section~4. Section~5 summarizes our main 
conclusions and comments on the role of temperature.
\section*{2. The role of band curvature in particle density fluctuations}
\label{sec2}
We consider a clean 1D liquid of interacting particles on a segment of length $L\rightarrow \infty$, 
which are described by the Hamiltonian [10]
\begin{equation}
H=K+U=\sum_{k,p}\varepsilon_{\alpha}(pk){\hat c}_{k,p}^{\dagger}{\hat c}_{k,p}+\frac{1}{2L}\sum_{k}V(k)\left[{\hat \rho}(k){\hat \rho}(-k)-{\hat N}\right]~.
\end{equation}
Here and in the following spin indices and sums over them are left implicit, since we are only 
interested in the total particle current density. In Eq.~(1) $p =\pm 1$ is the branch index, while the 
operators ${\hat c}_{k,p}^{\dagger}$  (${\hat c}_{k,p}$) create (annihilate) a particle of momentum $k$ and branch index $p$ and obey 
the canonical fermionic anticommutation relations $\{{\hat c}_{k,p}^{\dagger},{\hat c}_{k',p'}\}=\delta_{p,p'}\delta_{k,k'}$. The symmetric density 
operator ${\hat \rho}(k)$ is defined as
\begin{equation}
{\hat \rho}(k)=\sum_{p}{\hat \rho}_{p}(k)=\sum_{q,p}{\hat c}_{q-k,p}^{\dagger}{\hat c}_{q,p}
\end{equation}
and the single-particle spectrum $\varepsilon_{\alpha}(pk)$ is taken as
\begin{equation}
\varepsilon_{\alpha}(pk)=v_F(pk-k_F)+\frac{\alpha}{2 m}~(pk-k_F)^2~,
\end{equation}
with $k_F$ the Fermi momentum and $v_F=k_F/m$ ($\hbar = 1$). The linear Luttinger model (LLM) takes 
$\alpha=0$ in Eq.~(3), while the original Haldane spectrum is recovered by taking $\alpha=1$. 

We evaluate the collective particle-density excitations coming from the Hamiltonian (1) by 
the equation of motion approach, which was first introduced for the LLM by Metzner and Di 
Castro~[17]. The relevant commutators are
\begin{equation}
\left[{\hat \rho}_p(k),K\right]=\sum_q\left[\varepsilon_{\alpha}(pq)-\varepsilon_{\alpha}(p(q-k))\right]
~{\hat c}_{q-k,p}^{\dagger}{\hat c}_{q,p}
\end{equation}
and
\begin{equation}
\left[{\hat \rho}_p(k),U\right]=pk~\frac{V(k)}{2 \pi}~{\hat \rho}(k)~.
\end{equation}
We separately treat below the two cases $\alpha=0$ and $\alpha= 1$.
\subsection*{2.1 The LLM case ($\alpha=0$)}
For $\alpha=0$ Eq.~(4) yields $[{\hat \rho}_p(k),K]=p v_Fk{\hat \rho}_p(k)$ and hence $i \partial_t{\hat \rho}(k)=v_Fk{\hat \rho}_{\rm as}(k)$, where the asymmetric density operator is given by 
${\hat \rho}_{\rm as}(k)=\sum_{q,p}p{\hat c}_{q-k,p}^{\dagger}{\hat c}_{q,p}$. Using Eqs.~(4) and (5) to 
get the equation of motion for ${\hat \rho}_{\rm as}(k)$, 
one easily finds an undamped harmonic-oscillator equation for the density fluctuations, $\partial^2_t\hat{\rho}(k,t)+\Omega^2(k)\hat{\rho}(k,t)=0$ where
\begin{equation}
\Omega^2(k)=(v_Fk)^2\left[1+2V(k)/(\pi v_F)\right]~.
\end{equation}
This is the result that is commonly obtained by the bosonization technique for the Luttinger 
liquid (see {\it e.g.} Samokhin [9]). It has also been obtained for a 1D Coulomb fluid by the standard 
Bohm-Pines RPA [18] and, taking in this case $V(k)=2 e^2 K_0(kd)$ with $K_0(x)$ the modified 
Bessel function and $d$ the transverse width of the confinement, one recovers the well-known 
result $\Omega(k)\rightarrow (4 e^2 v_F/\pi)^{1/2}~k |\ln{(kd)}|^{1/2}$ for the plasma frequency at long wavelengths.

Starting from the undamped excitations described by Eq.~(6), Samokhin [9] has evaluated 
the spontaneous decay of one excitation into two at temperature $T=0$, as induced by the 
quadratic term in Eq.~(3) which plays the role of an effective interaction between the Tomonaga-Luttinger bosons. His calculation assumes a constant value for the interaction potential     
($U_0$, say) and yields from Eq.~(6) excitations having a sound-wave dispersion relation, $\Omega(k)=ck$ with $c=v_F\left[1+2U_0/(\pi v_F)\right]^{1/2}$. The imaginary part of the self-energy of the excitations at 
frequency $ck$ has the form $\gamma_k \propto k^2$ at $T=0$.

\subsection*{2.2 Current density in the case $\alpha=1$}
In the case $\alpha=1$ all linear terms cancel out inside the brackets in Eq.~(4). 
One finds
\begin{equation}
\left[\hat{\rho}_p(k),K\right]=k
\left[\sum_q\left(\frac{q}{m}\hat{c}^{\dagger}_{q-k,p},
\hat{c}_{q,p}\right)-\frac{k}{2m}\hat{\rho}_p(k)\right]
\end{equation}
and hence one obtains the continuity equation in the usual form,
\begin{equation}
i\partial_t\hat{\rho}(k,t)=k\hat{J}(k,t)~.
\end{equation}
Here, the current density $\hat{J}(k,t)$ has the standard expression for a fluid of Fermi particles, 
\begin{equation}
\hat{J}(k)=\sum_{q,p}\left(\frac{q}{m}\hat{c}^{\dagger}_{q-k,p}\hat{c}_{q,p}\right)-\frac{k}{2m}\hat{\rho}(k)=\sum_{q,p}\left(\frac{q}{m}{\hat c}^{\dagger}_{q-k/2,p} {\hat c}_{q+k/2,p}\right)~.
\end{equation}
(see {\it e.g.} Singwi and Tosi [19]). 

The equation of motion for the current density follows by taking the commutator of ${\hat J}(k)$ 
with the Hamiltonian~(1),
\begin{equation}
i\partial_t\hat{J}(k,t)=\frac{2k}{m}\hat{E}(k,t)+
\frac{1}{L}\sum_q\frac{q}{m}V(q)\hat{\rho}(k-q,t)\hat{\rho}(q,t)
\end{equation}
where the operator ${\hat E}(k)$ has been defined as
\begin{equation}
{\hat E}(k)=\sum_{q,p}\frac{q^2}{2m}\hat{c}^{\dagger}_{q-k/2,p}\hat{c}_{q+k/2,p}~.
\end{equation}
The equation of motion for particle density fluctuations follows from Eqs.~(8) and (10),
\begin{equation}
\partial^2_t {\hat \rho}(k,t)=-\frac{2k^2}{m}~{\hat E}(k,t)-\frac{1}{L}~\sum_q~\frac{kq}{m}~V(q){\hat \rho}(k-q,t){\hat \rho}(q,t)~.
\end{equation}		                                   
The RHS of Eq.~(12) contains both the restoring forces on density fluctuations and their 
damping by spontaneous decay and mutual scattering. The leading long-wavelength term for a 
1D Coulomb fluid is obtained in the RPA by preserving only the term $q=k$ in the summation 
over momenta and by dropping the kinetic term: one finds undamped oscillations at a frequency 
$[nk^2V(k)/m]^{1/2}$ with $n=N/L=2k_F/\pi$ being the electron density. The limiting value is again 
the plasma frequency $(4 e^2 v_F/\pi)^{1/2} k |\ln{(kd)}|^{1/2}$. 

In summary, we have seen that full account of the quadratic term in the single-particle 
spectrum in Eq.~(3) formally leads to the same equations of motion for particle number and 
longitudinal current density fluctuations as for Fermi liquids in higher dimensionalities. We 
proceed in the next section to transcend the RPA by calculating the exchange-correlation 
spectrum for current density fluctuations at long wavelength.
\section*{3. The exchange-correlation spectrum}
\label{sec3}
The exchange-correlation kernel  $f^{(L)}_{xc}(\omega)$ for longitudinal currents in a fermion fluid is a 
quantity of crucial importance for time-dependent density functional theory [20, 21]. 
It is defined as the long-wavelength limit of the function
\begin{equation}
f^{(L)}_{xc}(k,\omega)=
\frac{\omega^2}{k^2}\left[\frac{1}{\chi^0_L(k,\omega)+n/m}-\frac{1}{\chi_L(k,\omega)+n/m}\right]-V(k)~,
\end{equation}
where $\chi_L(k,\omega)$ is the current-current linear response function and $\chi^{0}_L(k,\omega)$ 
is the same function in the absence of interactions. 
Explicitly, the current response function is given by $\chi_L(k,\omega)=\langle\langle\hat{J}(k);\hat{J}(-k)\rangle\rangle_{\omega}$, according to the general definition of a response function in terms of unequal-time commutators:
\begin{equation}
\langle\langle\hat{A};\hat{B}\rangle\rangle_{\omega}\equiv-i\int_0^{\infty}dt~\langle[\hat{A}(t);\hat{B}(0)]\rangle~\exp{[i(\omega+i\varepsilon)t]}~.
\end{equation}      
Here ${\hat A}(t)$ is the operator in the Heisenberg representation, the brackets denote the ground-state 
expectation value, and $\varepsilon$ is a positive infinitesimal. 

The last term on the RHS of Eq.~(13) ensures that collective excitations do not contribute 
to the kernel at long wavelengths. In fact, to leading order in this limit the exchange-correlation 
spectrum is given by
\begin{equation}
{\rm Im}f^{(L)}_{xc}(\omega)=
\lim_{k\rightarrow 0}\frac{m^2\omega^2}{n^2k^2}{\rm Im}\tilde{\chi}_L(k,\omega)
\end{equation}
where $\tilde{\chi}_L(k,\omega)$ 
is the proper current response function. The real part of the kernel can then be 
obtained from the spectrum by means of the Kramers-Kronig relation [11]. 

Returning to the expression for $\chi_L(k,\omega)$ and following the treatment given by Nifos\`\i\ {\it et 
al.} [11], we find that the current response function satisfies the equation of motion
\begin{equation}
\omega^2\chi_L(k,\omega)=\langle[[\hat{J}(k),H],\hat{J}(-k)]\rangle-\langle\langle[\hat{J}(k),H];[\hat{J}(-k),H]\rangle\rangle_{\omega}
\end{equation}
and hence a lengthy but straightforward calculation yields
\begin{equation}
{\rm Im}\tilde{\chi}_L(k,\omega)=\frac{1}{m^2L^2\omega^4}\sum_{q,q'}
\Gamma(q,k)\Gamma(q',-k){\rm Im}\langle\langle\hat{J}(q)\hat{\rho}(-q);\hat{J}(q')\hat{\rho}(-q')\rangle\rangle_{\omega}+o(k^2)
\end{equation}
where
\begin{equation}
\Gamma(q,k)=q^2[V(\vert q+k\vert)-V(q)]-qkV(q).
\end{equation}
The four-point response function entering the RHS of the exact result in Eq.~(17) describes the 
spectrum of particle-hole multipair excitations in the interacting Fermi liquid, with a leading 
contribution coming from both direct and exchange two-pair processes. 

Following again Nifos\`\i {\it et al.} [11], the direct contribution to the two-pair channel can be 
extracted from Eq. (17) by the approximate decoupling
\begin{eqnarray}
{\rm Im}\langle\langle\hat{A}\hat{B};\hat{C}\hat{D}\rangle\rangle_{\omega}&\cong&-\int_0^{\omega}\frac{d\omega'}{\pi}[{\rm Im}\langle\langle\hat{A};\hat{C}\rangle\rangle_{\omega'}{\rm Im}\langle\langle\hat{B};\hat{D}\rangle\rangle_{\omega-\omega'}+\nonumber\\
& &+{\rm Im}\langle\langle\hat{A};\hat{D}\rangle\rangle_{\omega'}{\rm Im}\langle\langle\hat{B};\hat{C}\rangle\rangle_{\omega-\omega'}]~.
\end{eqnarray}
Such a decoupling approximation should be corrected for exchange terms, which are known to 
be very small at low frequency both on the basis of a general argument [11] and of detailed 
calculations in higher dimensionalities [13]. However, the exchange corrections reduce the value 
of ${\rm Im} f^{(L)}_{xc}(\omega)$ by a factor $1/2$ in the high-frequency limit, which was studied in various dimensionalities by Glick and Long [22] and by Holas and Singwi [23]. 

By using Eq.~(19) in Eq.~(17) we obtain from Eq.~(15) the exchange-correlation spectrum 
at long-wavelengths in the two-pair approximation in the form
\begin{equation}
{\rm Im}f^{(L)}_{xc}(\omega)=-(2 \pi^2 n^2)^{-1}~a_L~g_x(\omega)~\int_0^{\omega}d\omega'\int^{\infty}_{-\infty}dq~\vert V(q)\vert^2~ {\rm Im}\chi(q,\omega')~{\rm Im}\chi(q,\omega-\omega').
\end{equation}
Here, $a_L$ is a numerical coefficient to be obtained from the value of the kernel in the high-frequency limit, and $g_x(\omega)$ is a factor accounting for pair exchange processes which interpolates 
between $1$ at low $\omega$ and $0.5$ at high $\omega$ on their energy scale (twice the Fermi energy). In deriving the expression in Eq.~(20) we have also made use of the exact relation
\begin{equation}
\chi_L(k,\omega)+n/m=\omega^2\chi(k,\omega)/k^2
\end{equation}
holding between $\chi_L(k,\omega)$ and the density response function $\chi(k,\omega)$. 

At the level of second-order perturbation theory, the functions $\chi(k,\omega)$ in the integrand in 
Eq.~(20) should be replaced by their equivalents for the non-interacting Fermi fluid ($\chi_0(k,\omega)$, say). In this case the integral diverges when $V(k)$ is taken as the bare Coulomb interaction (the 
same divergence is in fact present for electron fluids in higher dimensionalities [11 - 13]). To 
account for all higher-order terms in an approximate way, one may draw on the analogy of the 
theory with the standard treatment of transport phenomena {\it via}  a collision integral and Fermi's 
golden rule. This suggests that in the integrand in Eq.~(20) one may replace $\chi(k,\omega)$ by $\chi_0(k,\omega)$ provided that at the same time the bare potential is replaced by an effective scattering potential. 
The use of Thomas-Fermi screening, or even the replacement of $V(k)$ by a constant, are the 
simplest approximations that one may invoke for actual numerical calculations. 

In summary, the expression that we propose for the exchange-correlation kernel in the 
two-pair approximation reads
\begin{equation}
{\rm Im}f^{(L)}_{xc}(\omega)=-(2 \pi^2 n^2)^{-1}~a_L~g_x(\omega)~\int_0^{\omega}d\omega'
\int^{\infty}_{-\infty}dq\vert V_{sc}(q)\vert^2{\rm Im}\chi_0(q,\omega'){\rm Im}\chi_0(q,\omega-\omega')~.
\end{equation}
Here, $V_{sc}(k)$ is a screened Coulomb potential, that in the next section we shall take as a constant $U_0$ to make connection with the experiments of Go${\tilde {\rm n}}$i {\it et al.}~[2] on the one hand and with the damping of excitations in a Luttinger liquid calculated by Samokhin [9] on the other.
\section*{4. Lifetime of excitations}
\label{sec4}
From the work of Holas and Singwi~[23] we find that in a 1D Fermi liquid the high-frequency behaviour of the spectrum of the proper density response function from two-pair 
processes is given by
\begin{equation}
{\rm Im}\tilde{\chi}(k,\omega)\rightarrow -\frac{m^3 k_F^2 k^4}{2\pi^2(m\omega)^{9/2}}U_0^2~,
\end{equation}
in the case of constant effective interactions. Equations~(15) and (21) then yield
\begin{equation}
{\rm Im}f_{xc}(\omega)\rightarrow -\frac{m^{1/2}U_0^2}{8\omega^{1/2}}~.
\end{equation}      
The same asymptotic result is obtained from Eq.~(22) provided  $g_x(\omega) \rightarrow 1/2$ and $a_L=1/2$. We have made use of the expression
\begin{equation}
{\rm Im}\chi_0(k,\omega)=-\frac{m}{k}~\vartheta\left(\left\vert\frac{k}{2k_F}+
\frac{m\omega}{k_Fk}\right\vert-1\right)\vartheta\left(1-\left\vert\frac{k}{2k_F}-\frac{m\omega}{k_Fk}\right\vert\right),
\end{equation} 
for the spectrum of the 1D ideal Fermi gas, with $\vartheta(x)$ being the Heaviside step function. 

We are now ready to evaluate the damping of collective excitations in the 1D Fermi liquid, 
on the assumption that their dispersion relation is of the acoustic type. We write $\omega_k=ck-i\gamma_k$ and for $|\gamma_k|\ll ck$ we have [11]
\begin{equation}
\gamma_k=-\frac{ck}{2U_0}~{\rm Im}f^{(L)}_{xc}(ck)~.
\end{equation}
Equation~(25) is again used to evaluate the integral in Eq.~(22) at low frequency, with the result 
\begin{equation}
\gamma_k=\frac{m^3c^3}{64k_F^5}~U_0k^3. 
\end{equation}  
In obtaining this expression we have taken $g_x(\omega)= 1$ at low frequency.

\section*{5. Concluding remarks}
\label{sec5}
In summary, we have calculated the decay of collective density excitations in a quasi-1D 
Fermi liquid from the process of decay into pairs of particle-hole excitations. We have assumed 
a short-range (contact) effective interaction and a linear dispersion relation. We have argued, 
however, that our final result in Eq.~(27) is also approximately applicable to a quasi-1D Fermi 
liquid of electrons in semiconductor quantum wires, in a range of wave number and wire 
thickness where an approximately linear dispersion relation is observed for the plasmon in 
inelastic scattering experiments. 

We have found that within this model the dependence of the decay rate on wave number is 
of the type $\gamma_k \propto k^3$ at $T=0$. The temperature dependence of the lifetime can be calculated by 
means of the fluctuation-dissipation theorem [14] and one easily finds $\gamma_k \propto T k^2$ in the regime $k_B T \gg \hbar c k$. Therefore, the behaviours of the decay rate within a Fermi-liquid model are clearly distinct from those predicted for a Luttinger liquid. For the latter Samokhin [9] has reported $\gamma_k \propto k^2$ at $T=0$ and $\gamma_k(T) \propto T^{1/2} k^{3/2}$ at high temperature.
\vspace{0.5 cm}

\noindent {\bf Aknowledgements}
\vspace{0.2 cm}

This work was partially supported by MIUR within the PRIN2001 Initiative and by 
INFM through the PRA-Photonmatter. Two of us (M.P. and M.P.T.) acknowledge useful 
contributions from Dr A. Gama Goicochea in the initial part of this work. The hospitality of the 
Condensed Matter Group of the Abdus Salam ICTP during the final stages of this work is 
gratefully acknowledged.
\newpage
     
\noindent {\bf References}
\vspace{0.2 cm}

[1] H. J. Schulz, G. Cuniberti and P. Pieri, in "Field Theories for Low-Dimensional 
	  Condensed Matter Systems", 
	  
	  \hspace{0.35 cm} eds. G. Morandi, P. Sodano, A. Tagliacozzo and V. 
	  Tognetti (Springer, Berlin 2000) p. 9, and references given 
	  
	  \hspace{0.35 cm} therein. \\
	
[2] A. R. Go${\tilde {\rm n}}$i, A. Pinczuk, J. S. Weiner, J. M. Calleja, B. S. Dennis, L. N. Pfeiffer and K. W. West, 

\hspace{0.35 cm} Phys. Rev. Lett. 67 (1991) 3298. \\
	
[3]	D. W. Wang, A. J. Millis and S. Das Sarma, Phys. Rev. Lett. 85 (2000) 4570.\\

[4]	B. Y.-K. Hu and S. Das Sarma, Phys. Rev. B 48 (1993) 5469.\\

[5]	Q. P. Li and S. Das Sarma, Phys. Rev. B 43 (1991) 11768.\\

[6]	H. Monien, M. Linn and N. Elstner, Phys. Rev. A 58 (1998) R3395.\\

[7]	W. Wonneberger, Phys. Rev. A 63 (2001) 063607.\\

[8]	A. Recati, P. O. Fedichev, W. Zwerger and P. Zoller, cond-mat/0206424.\\

[9]	K. V. Samokhin, J. Phys.: Condens. Matter 10 (1998) L533.\\

[10]	F. D. M. Haldane, J. Phys. C 14 (1981) 2585.\\

[11]	R. Nifos\`\i, S. Conti and M. P. Tosi, Phys. Rev. B 58 (1998) 12758.\\

[12]	S. Conti and G. Vignale, Phys. Rev. B 60 (1999) 7966.\\

[13]	Z. Qian and G. Vignale, Phys. Rev. B 65 (2002) 235121.\\

[14] S. Conti, A. Minguzzi and M. P. Tosi, Phys. Lett. A 250 (1998) 177.\\

[15]	A. Minguzzi and M. P. Tosi, Phys. Lett. A 262 (1999) 361.\\

[16] B. Tanatar, Phys. Rev. B 51 (1995) 14410.\\

[17] W. Metzner and C. Di Castro, Phys. Rev. B 47 (1993) 16107.\\

[18]	Q. P. Li, S. Das Sarma and R. Joynt, Phys. Rev. B 45 (1992) 13713.\\

[19] K. S. Singwi and M. P. Tosi, in "Solid State Physics" vol. 36, eds. H. Ehrenreich, F. 
	Seitz and D. Turnbull 
	
	\hspace{0.50 cm} (Academic, New York 1981) p. 177.\\
	
[20]	G. Vignale and W. Kohn, Phys. Rev. Lett. 77 (1996) 2037.\\

[21] G. Vignale and W. Kohn, in "Electronic Density Functional Theory", eds. J. Dobson, M. 
	P. Das and G. Vignale 
	
	\hspace{0.50 cm} (Plenum, New York 1997).\\
	
[22]	A. J. Glick and W. F. Long, Phys. Rev. B 4 (1971) 3455.\\

[23] A. Holas and K. S. Singwi, Phys. Rev. B 40 (1989) 158.\\
\end{document}